\documentclass[12pt,a4paper]{conference}

\usepackage{fancyhdr,epsfig}
\usepackage{graphicx,amsmath,amssymb,cite}
\usepackage{multind}

\newcommand{\beq}{\begin{equation}}
\newcommand{\eeq}[1]{\label{#1}\end{equation}}
\newcommand{\eeqn}{\end{equation}}

\newcommand{\beqa}{\begin{eqnarray}}
\newcommand{\eeqa}[1]{\label{#1}\end{eqnarray}}
\newcommand{\eeqan}{\end{eqnarray}}

\let\bar=\overbar

\newcommand{\Dslash}{\not{\hbox{\kern-4pt $D$}}}
\newcommand{\dslash}{\not{\hbox{\kern-2pt $\del$}}}

\newcommand{\msb}{{\bar{\ssstyle M \kern -1pt S}}}

\begin{document}

\Chapter{Hadrons in Medium -- \\Theory meets Experiment}
           {Hadrons in Medium}{U. Mosel}
\vspace{-6 cm}\includegraphics[width=6 cm]{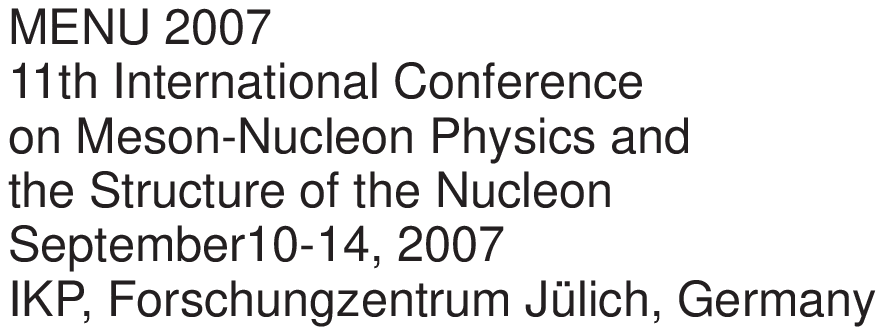}
\vspace{4 cm}

\addcontentsline{toc}{chapter}{{\it U. Mosel}} \label{authorStart}

\begin{raggedright}

{\it U.\ Mosel \footnote{mosel@physik.uni-giessen.de}}\index{author}{Mosel, U.}\\
Institut f\"ur Theoretische Physik\\
Universit\"at Giessen\\
Giessen, Germany, D-35392
\bigskip\bigskip

\end{raggedright}

\begin{center}
\textbf{Abstract}
\end{center}
In this talk I give a short review of theoretical results on the
properties of hadrons in cold, equilibrium nuclear matter. I then discuss
the observable consequences of any changes of these properties inside the
medium in actual experiments. I demonstrate that any experimental
verification of in-medium effects requires a state-of-the-art treatment of
the reaction dynamics and, in particular, also the final state
interactions.

\section{Introduction}

The interest in in-medium properties of hadrons has been growing over the
last decade because of a possible connection with broken symmetries of QCD
and their partial restoration inside nuclear matter. Already two decades
ago Bernard and Meissner predicted on the basis of the NJL model that the scalar strength should drop considerably inside nuclei whereas the vector mesons were only little affected in that approach \cite{BM}. Somewhat later there were theoretical predictions that masses of vector
mesons should generally decrease in medium as a function of density due to
a partial restoration of chiral symmetry \cite{Brown-Rho,HL}.
In addition, there existed well-worked out predictions that the scalar
$2\pi$ strength should decrease in medium and that the vector meson masses
should drop \cite{Kuni,Klingl}. All of these calculations were performed
for idealized situations (infinite cold nuclear matter at rest) and little
attention was paid to the actual observability of these predicted changes.
At the same time experiments (CERES, TAPS) seemed to show the predicted
behavior. The CERES results indicated a significant broadening of the
$\rho$ meson in medium \cite{Ceres}, whereas the TAPS results on the
$2\pi$ strength exhibited the predicted lowering of the $\sigma$ strength
inside nuclei \cite{TAPS2pi}. Most recently, the CBELSA/TAPS experiment
has also obtained an indication for a lowering of the $\omega$ meson mass
in nuclei \cite{Trnka}. Other interesting data in this context have been
obtained by groups at JLAB \cite{g7}, KEK \cite{KEK} and CERN \cite{NA60}.

Motivated by these developments we have concentrated our theoretical work
on in-medium properties along two different lines. First, we have
performed state-of-the-art calculations of vector meson spectral functions
in cold nuclear matter. Second, we have followed closely the CBELSA/TAPS
experiment and have performed various feasibility studies and analyses of
this experiment searching for in-medium changes of the $\omega$ meson in
medium. In a third step we have also analyzed the lowering of scalar
strength in nuclei observed by the TAPS experiment.

\section{In-Medium Properties of Vector Mesons}

On the first aspect we have initially finished a major calculation
on the in-medium properties of the $\pi, \rho$ and $\eta$ mesons
\cite{Post1}. In this work we have generated the in-medium
selfenergies of these mesons by nucleon-hole and resonance-hole
excitations which in turn are affected by the changed in-medium
properties of the mesons. This self-consistency problem has been
solved here for the first time. Special care was taken to respect
the analyticity of the spectral functions and to take into account
effects from short-range correlations both for positive and
negative parity states.

\begin{figure}
\centerline{\epsfig{file=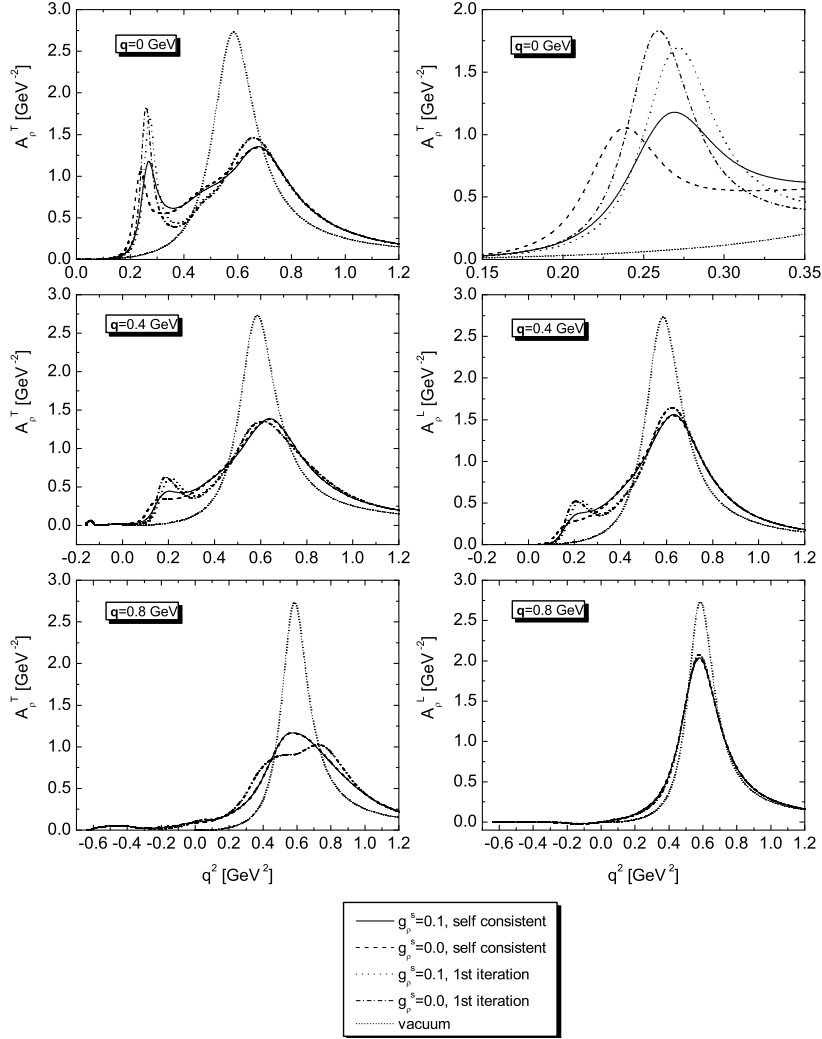,height=14cm}}
\caption{Spectral function of the $\rho$ meson in nuclear matter
at density $\rho_0$ for various momenta indicated in the figure.
The left column shows the transverse spectral function, the right
column that of longitudinally polarized $\rho$ mesons. The thin
dotted line in each figure is the vacuum spectral function, the
other curves give the effect of selfconsistency and short-range
correlations (from \cite{Post1}).} \label{rhospect}
\end{figure}

Our model has been shown to produce sensible results for pion and
$\Delta$ dynamics in nuclear matter, as a test. For the $\rho$
meson we find a strong interplay with the D13(1520), which moves
spectral strength of the $\rho$ spectrum to smaller invariant
masses and simultaneously leads to a broadening of the baryon
resonance. The strong interplay between the $\rho$ meson and the
D13(1520)-nucleon hole excitation leads to a dominant lower hump
in the $\rho$ spectral function also in this relativistic and
selfconsistent calculation; it confirms our earlier result
obtained in a more simplified approach. Whereas the longitudinal
component of the $\rho$ meson only broadens somewhat, the
transverse component shows a major distortion which evolves as a
function of the $\rho$ momentum (see Fig. \ref{rhospect}). At the
same time, the D13(1520) resonance broadens considerably due to
the opening of phase-space for $\rho$-decay. For the $\eta$ meson
the optical potential resulting from our model is rather
attractive whereas the in-medium modifications of the S11(1535)
are found to be quite small.

These studies also allow us to assess the validity range of the
often used low-density approximation. We find that this depends
very much on the special couplings involved and thus varies from
meson to meson. Whereas for the $\eta$ meson the validity ranges
up to a density of $\rho_0$, for the $\rho$ meson it already
breaks down at about 0.3 $\rho_0$. This may serve as a warning
sign for many in-medium calculations that use the low-density
approximation without any further proof of its reliability.

\begin{figure}
\centerline{\epsfig{file=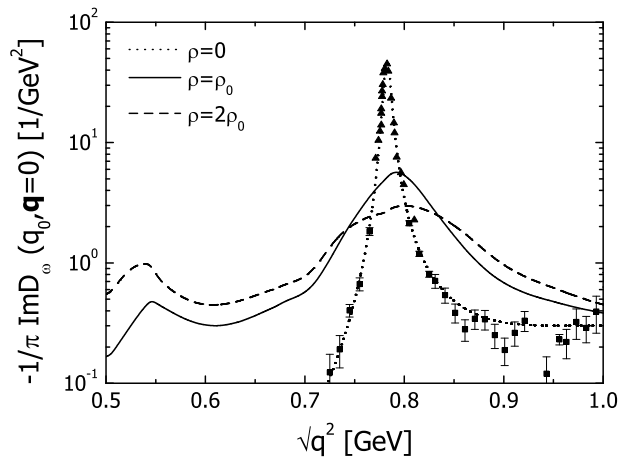,height=6cm}}
 \caption{Spectral function of the $\omega$ meson in nuclear matter at rest,
at densities $0$, $\rho_0$ and 2 $\rho_0$ (from
\cite{Muehlich:2006nn}).} \label{omspectr}
\end{figure}

Bearing this in mind we have recently also performed a calculation of the
selfenergy of the $\omega$ meson in medium. This calculation is based on a
unitary coupled channel analysis of all existing $\pi N$ and $\gamma N$
data up to an invariant mass of 2 GeV \cite{Muehlich:2006nn}. The coupled
channel character of this calculation is of utmost importance here because
it is the only way to include experimental constraints on the $2\pi$ decay
channel that was found to be dominant in \cite{Klingl}. This analysis and
thus also the selfenergy of the $\omega$ meson extracted from the $\omega
N$ scattering amplitude gives a broadening of about 60 MeV at $\rho_0$ and
a small upward shift of the peak mass. In addition, due to a nonzero
coupling of the $\omega$ to the S11(1535) resonance the $\omega$ spectral
function exhibits a small peak at a mass of around 550 MeV. This
calculation gives, for the first time, the $\omega$ selfenergy also for
nonzero momenta (which corresponds to the experimental situation) and it
takes the experimental constraints on the important $2\pi$ channel into
account because it is based on a unitary $K$-matrix analysis of 'real'
data. The result of this calculation is shown in Fig. \ref{omspectr}.
For vanishing $\omega$ momentum this result qualitatively agrees
with that of \cite{Lutz} in that it yields a lower mass structure
in the spectral function at an invariant mass of 500 - 600 MeV and
only a very small shift of the main peak; the latter is in
contrast to the results of \cite{Klingl} (for a recent discussion
of the results obtained in \cite{Klingl} see \cite{Eichst}) and of
\cite{Thomas}. The latter was based on a relativistic mean-field
model and does not contain any dispersive effects.

Our second line of approach to the problem of in-medium
selfenergies has concentrated on an analysis of the recent
CBELSA/TAPS data \cite{Trnka} Since the experiment looks for the
channel $\gamma + A \to A^* + \omega \to A^* + \gamma + \pi^0$ it
is mandatory to control the effects of final state interactions on
the $\pi^0$ in a quantitative way. The only method available for
this is that of coupled channel semiclassical transport
calculations which -- as we had shown earlier in extensive work --
can give a consistent description of many experimental phenomena,
both in heavy-ion \cite{Lari} as well as in nucleon-, pion-
\cite{Buss}, photon- \cite{Buss-Leitner} and neutrino-induced
reactions \cite{Leitner}. For any reaction on nuclei with hadrons
in the final state a state-of-the-art transport calculation of the
final state interactions is an indispensible part of the theory.
We have, therefore, spent significant effort on developing a new
code, dubbed 'GiBUU', for the transport calculations. This code is
written in object-oriented FORTRAN 95/2003 and incorporates all
the experience we have gained with earlier numerical
implementations at Giessen over the last 20 years \cite{GiBUU}.

With this method we have first analysed both results on the
experimental determination of the nuclear transparency ratio for
$\phi$ mesons \cite{Muhlich:phi}, measured by a group at SPRING8.
This transparency gives directly the imaginary part of the meson's
selfenergy in medium; using a low-density approximation one can
then extract the inelastic $\omega N$ cross section. In this way
an an unexpectedly large inelastic cross section for $\phi N$
interactions was extracted. We have found that indeed cross
sections about a factor 3 larger than those theoretically expected
are needed to explain the mentioned data, in line with a simple
Glauber analysis by the SPRING8 group.

For $\omega$ mesons the CBELSA/TAPS collaboration has measured the
nuclear transparency \cite{Metag}. We have shown that our
calculations reproduce the measured attenuation quite well if --
similar to the $\phi N$ case -- the inelastic cross section is
increased by about a factor of 3 beyond earlier theoretical
expectations. A fit to the data can actually also determine the
momentum-dependence of this cross section \cite{MuehlichDiss1}.

\begin{figure}[htb]
\centerline{\epsfig{file=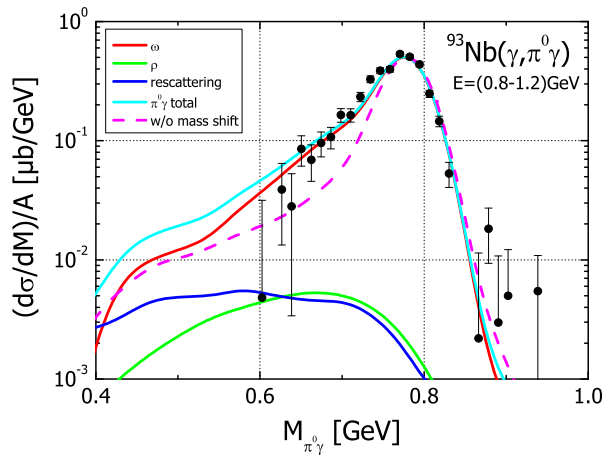,height=6cm}}
\caption{Comparison of CBELSA/TAPS data on $\omega$ production on
different targets in the energy range $E_\gamma = 0.8 - 1.2$ GeV
\cite{Trnka}. The dashed line shows a comparison of the GiBUU
transport calculation with free $\omega$ meson spectral functions
(using the experimental mass resolution) and the solid line at the
top gives the results of a calculation with mass-shift and
collisional broadening. The solid line at the bottom gives the
contribution of rescattered pions to the reconstructed spectral
function and the grey line at the bottom gives the contribution of
the decay channel $\rho \to \pi^0 \gamma$. The top  curve shows
the sum of all contributions (from \cite{MuehlichDiss1}).}
\label{CBTAPS-omega}
\end{figure}

A major effort has gone into an analysis of the $\omega$
photo-production experiment at CBELSA/TAPS. Fig.
\ref{CBTAPS-omega} shows the result of such an analysis together
with the data of the CBELSA/TAPS collaboration.
Our simulations give a full event analysis and thus allow to calculate
also background contributions on the same footing as the actual signal.
They also allow insight into the effects of rescattering of the pions
produced in the decay of the $\omega$ meson and have suggested a method to
suppress the rescattered pion background that has actually been adopted by
the experimental group. The result of this analysis is that the data can
be explained if a lowering of the $\omega$ meson mass in medium by about
16 \% is assumed together with the appropriate collisional broadening.

A problem in this context is that the experiment does not
determine the spectral function of the $\omega$ meson itself.
Instead, we have noted that the result of the experimental
analysis is the product of the primary production cross section
with the spectral function and the partial decay probability into
the channel under study ($\pi^0 \gamma$ here). If the first and
the latter depend strongly on the invariant mass itself, as it
does for the CBELSA/TAPS experiment, then significant distortions
of the spectral function may arise. This is a topic under
intensive study by us presently \cite{MosEr}

\section{In-Medium Properties of Scalar Mesons}

Finally, we have analyzed the TAPS data on $2\pi^0$ photoproduction on
nuclei. The motivation for this experiment was a prediction that -- due to
chiral symmetry restoration in nuclei -- the scalar strength of the
$\sigma$ meson should be lowered in nuclei \cite{Kuni}. The TAPS
collaboration had indeed initially seen an effect as predicted in the
$2\pi^0$ data, whereas a comparison measurement in the $\pi^0 \pi^{\pm}$
channel did not seem to show such an effect \cite{TAPS2pi}. Various
explanations for these findings have been advanced by the Valencia group
\cite{Oset2pi} and by a group in Lyon \cite{Chanfray} in terms of chiral
symmetry restoration or $\pi - \pi$ correlations in nuclei, based on a
chiral effective field theory model.

\begin{figure}[htb]
\centerline{\epsfig{file=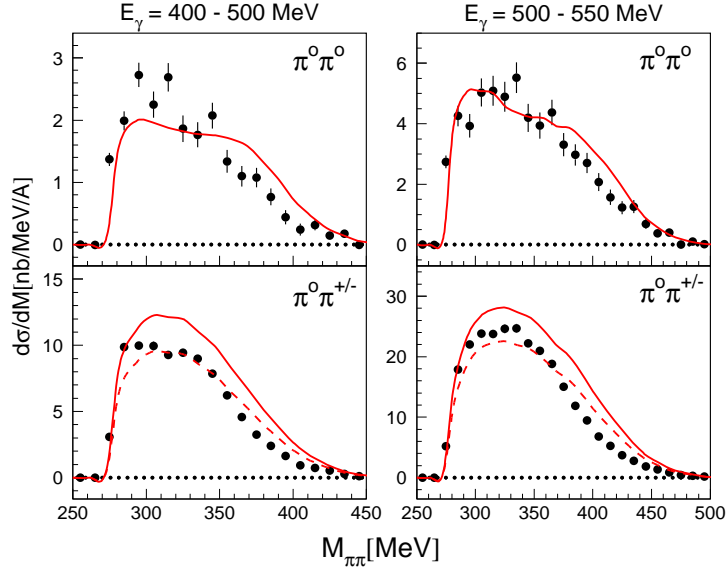,height=8cm}} \caption{Data of the TAPS
collaboration (Bloch et al.) for 2$\pi^0$ photoproduction on a $^{40}Ca$
target for two different photon energies. The solid curve gives the result
of a GiBUU calculation \cite{Buss1}, the dashed curves in the semicharged
2$\pi$ channel are normalized to the data (see text). Data from
\cite{Bloch}.} \label{2pi}
\end{figure}

None of these calculations, however, did look into the simplest
possible explanation of the observed effects in terms of mundane
pion rescattering. We have, therefore, performed such calculations
\cite{Buss} using the GiBUU method which is ideally suited for
this task. These calculations, which did not contain any effects
connected with $\pi \pi$ correlations, could reproduce the
observed effect for the $\pi^0 \pi^0$ channel, but they also
predicted a similar effect in the semi-charged channel where it
had not been seen experimentally. However, a more recent analysis
with higher statistics by Bloch et al. \cite{Bloch}  yielded a
result for both charge channels that is in perfect agreement with
our calculations (see Fig. \ref{2pi}; the dashed lines in this
figure are normalized in height to the data, this normalization
reflects uncertainties in the elementary cross sections). In
particular the yield in the semi-charged channel is strongly
influenced by a coupled-channel effect, the charge transfer in
$\pi N$ interactions; Glauber based absorption models miss this
contribution. This illustrates that a very sophisticated treatment
of final state interactions is absolutely mandatory when looking
for more 'exotic' effects in nuclei. We conclude that any analysis
of the $2\pi^0$ data with respect to a lowering of the scalar
strength in nuclei has to take the pion rescattering effects into
account. Present day's data are all consistent with simple
rescattering.

\section{Summary}
The main message we have learned from the studies reported here is that it
is important to calculate not only in-medium properties under idealized
conditions (static, uniform matter in equilibrium), but to also explore
the influence of these properties on actual observables. The spectral
function itself, which contains the information on in-medium selfenergies,
in particular in-medium masses and widths, is not directly observable.
Instead, both the creation of the studied hadron as well as its decay
influence the observables as much as the spectral function itself and thus
have to be under good control. The same holds for the final state
interactions on hadronic decay products. Here a state-of-the-art treatment
of final state interactions is mandatory. There is now general agreement
on the amount of collisional broadening of vector mesons in medium, but
the verification of an actual mass-shift still requires more work, both
theoretical and experimental.

\end{document}